\documentclass[a4paper,fleqn]{cas-sc}

\usepackage{subcaption}
\usepackage{multirow}
\usepackage{booktabs}      
\usepackage{siunitx}      
\usepackage[numbers]{natbib}

\def\tsc#1{\csdef{#1}{\textsc{\lowercase{#1}}\xspace}}
\tsc{WGM}
\tsc{QE}

\begin{document}
	\let\WriteBookmarks\relax
	\def\floatpagepagefraction{1}
	\def\textpagefraction{.001}

	\title [mode = title]{Image Encryption via Data‑Identified Discrete Chaotic Maps}  

	\author[1]{\textcolor{black}{Wenyuan Li}}\credit{Conceptualization, Methodology, Writing-original draft}
	\author[1,2]{\textcolor{black}{Xiao-Yun Wang}}\corref{cor1}\ead{xywang@lut.edu.cn}
	\credit{Supervision, Funding acquisition, Writing-review \& editing}
	\author[1,2]{\textcolor{black}{Zhigang Zhu}}
	\credit{Project administration, Resources}
	\author[1]{\textcolor{black}{Xiaofeng Zhang}}
	\credit{Investigation, Software}
	\author[1]{\textcolor{black}{Li Zhang}}
	\credit{Data curation, Formal analysis}
	\affiliation[1]{organization={Lanzhou University of Technology},
		department={Department of Physics},
		city={Lanzhou},
		postcode={730050},
		country={China}}
	
	\affiliation[2]{organization={Lanzhou University},
		addressline={Lanzhou Center for Theoretical Physics, 
			Key Laboratory of Theoretical Physics of Gansu Province, 
			and Key Laboratory of Quantum Theory and Applications of MoE},
		city={Lanzhou},
		postcode={730000},
		state={Gansu},
		country={China}}

	\cortext[cor1]{Corresponding author}

	\begin{abstract}
	In this work, we propose a data-driven image encryption framework that identifies chaotic maps directly from data via the SINDy-PI algorithm. Unlike conventional encryption schemes relying on predefined maps, our method learns the full explicit dynamics—including cross-terms and higher-order nonlinearities—from observational data. The validity of this approach is verified on three distinct chaotic systems: the Hénon map, the three‑dimensional logistic map, and the piecewise‑linear Lozi map, demonstrating its generality. The encryption key consists solely of initial conditions; the map structure itself becomes data‑dependent, introducing an extra layer of security. Moreover, even when the initial conditions are fixed, different training data (e.g., with a tiny noise seed) lead to slightly different maps, which produce completely different ciphertexts (NPCR $\approx 99.6\%$, UACI $\approx 33.5\%$). Numerical experiments on the Hénon system show near-ideal information entropy ($\approx 8$~bits), negligible inter-pixel correlation, and extreme sensitivity to initial conditions: a perturbation of $10^{-16}$ causes total decryption failure. The scheme resists both differential and statistical attacks, with NPCR and UACI values matching theoretical ideals. Our results establish a new paradigm for chaos-based cryptography beyond fixed maps.
	\end{abstract}

	\begin{highlights}
		\item First adaptation of SINDy-PI to discrete chaotic maps.
		\item Data-dependent map acts as an implicit encryption key.
		\item Validated on Hénon, 3D logistic, and Lozi maps.
		\item Open-source code for reproducible encryption.
	\end{highlights}
	
	\begin{keywords}
	Image encryption\sep SINDy-PI\sep Discrete chaotic map\sep
	\end{keywords}

	\maketitle
	
	\section{Introduction}\label{sec1}
	Over the past few decades, with the rapid development of multimedia technology, digital image has become one of the main information carriers in network interaction, and has been widely used in important fields such as military system, government agencies, financial system and health care. However, the security of images in the process of transmission and storage has become increasingly prominent and has received widespread attention. Ensuring the security of digital images during transmission and storage, while preventing unauthorized interception, tampering, or duplication, remains a critical challenge in cryptography\cite{Saberikamarposhti2024,Ozkaynak2018}.
	
	At present, significant progress has been made in the field of cryptography, and researchers have designed various encryption algorithms through various methods. Among them, the research of encryption algorithm based on chaos theory occupies an important position. The principle of this method is as follows : the initial values and parameters of the chaotic system are used as keys, and a series of real sequences are generated by using these keys and quantized into integer sequences. Then the chaotic integer sequence is reversibly interacted with data such as image, text or audio to achieve the purpose of encryption \cite{Zhang2025}.
	
	Chaotic systems (primarily discrete-time systems) play a significant role in the study of cryptographic algorithms\cite{Zhou2023}. Researchers have designed various algorithms by combining different chaotic systems. For example, the three-dimensional logistic map can be used to propose a novel image encryption scheme that differs from previous studies\cite{Qian2021}, and it can also be applied to text encryption based on research into chaotic image encryption algorithms\cite{Lawnik2022}. Additionally, the two-dimensional discrete Hénon map has been utilized to design new image encryption algorithms\cite{Ping2015}.
	
	Existing chaos-based encryption schemes predominantly rely on well-known systems, such as the Hénon map\cite{Ping2015} or three-dimensional logistic maps\cite{Qian2021,Khan2023}, or their variants. The security performance of algorithms designed based on these systems remains a concern, and low security continues to be a major challenge. If a completely new chaotic system could be proposed and applied to research in cryptography and related fields, it would not only provide researchers with more options for algorithm design but also offer new insights for overcoming the current issue of low algorithm security. However, due to the complexity of chaotic systems and their chaotic properties, proposing a completely new chaotic system is no easy task. Therefore, finding an efficient and relatively straightforward method to achieve the goal of designing new chaotic encryption algorithms is particularly important\cite{Nair2024}.
	
	The SINDy algorithm is a method for discovering dynamic system models from data\cite{Brunton2016}. Researchers have verified SINDy’s ability to accurately capture transient/steady-state behavior, noise effects, and model structure\cite{Pandey2024}, and it can even be extended to identify delayed differential equations\cite{Sandoz2023,EscobarRuiz2024}. Currently, building upon the SINDy algorithm, researchers have developed a novel variant, the SINDy-PI algorithm\cite{Kaheman2020}, which can identify the correct differential equations from limited data contaminated by noise. When identifying nonlinear systems, the algorithm independently identifies each equation comprising the system’s equation set. Since each equation consists of specific functional terms, as long as the preselected functions include these terms, the algorithm can identify the system’s functional terms and their coefficients through methods such as sparse regression, and combine them into the system equation to be identified. Consequently, when capturing unknown chaotic systems, if we can discretize the chaotic system into multivariable dynamical equations, this algorithm can be used to capture and reconstruct the unknown chaotic equations from real-world chaotic state data.
		
	Previously, the SINDy-PI algorithm was applied only to time-continuous systems. Given its power in identifying implicit dynamics, this paper extends it to discrete chaotic systems for the first time and proposes a data-driven adaptive image encryption framework. The core idea is to use SINDy-PI to automatically discover the complete explicit equation of a chaotic map from observational data (exemplified by the Hénon map, but possibly including cross-terms and higher-order nonlinearities). The approach is further validated on two other systems (3D logistic map\cite{Qian2021} and 2D Lozi maps\cite{Misiurewicz2016}), confirming its generality. The encryption key consists solely of the initial state $(x_0,y_0)$; the map itself is data‑dependent and not transmitted as a key. Importantly, even with fixed initial conditions, different training data lead to slightly different maps, producing completely different ciphertexts (NPCR $\approx99.6\%$, UACI $\approx33.5\%$). This data‑dependent nature acts as an implicit key, adding an extra security layer beyond the conventional key space. Compared to fixed‑map schemes, the method achieves extreme sensitivity to initial values ($10^{-16}$ perturbation causes decryption failure), near‑ideal information entropy ($\approx8$ bits), negligible inter-pixel correlation, and full compliance with NPCR/UACI benchmarks, effectively resisting reverse engineering and known-plaintext attacks.
		
	Taking the Hénon map as an example, this paper fully implements the entire process of “data acquisition → system identification → chaotic map construction → image encryption” and conducts a systematic security analysis\cite{Tuli2022}. Experimental results indicate that the proposed algorithm achieves or approaches theoretical optimal values in terms of information entropy, inter-pixel correlation, key sensitivity, and resistance to differential attacks, thereby validating the effectiveness and versatility of this framework. The structure of this paper is as follows: Sec.~\ref{sec2} introduces the chaotic system and algorithmic model used; Sec.~\ref{sec3} briefly describes the modifications to the SINDy-PI algorithm and its discrete adaptation; Sec.~\ref{sec4} details the image encryption scheme; Sec.~\ref{sec5} presents the experimental results and security analysis; Sec.~\ref{sec6} summarizes the paper and outlines future work.

	\section{System and Algorithm Model}
	\label{sec2}
	\subsection{The Classic Hénon Map}
	\label{sec2.1}
\begin{figure}[h]
		\centering
		\includegraphics[width=0.5\textwidth]{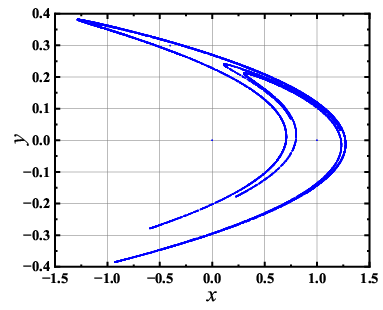}
		\caption{Two-dimensional Hénon map strange attractor}\label{fig:1}
	\end{figure}

	In 1976, inspired by Pomeau’s numerical results on the Lorenz system, the French mathematician Hénon simulated the system using three mapping chains of the $(x,y)$ plane itself and, by adjusting the parameters, derived the classic Hénon map. This is a two-dimensional mapping that, as a simple high-dimensional chaotic map, exhibits excellent nonlinear dynamical characteristics\cite{Ping2015}.The mathematical expression for this system is given by Eq.~\ref{eq:1}:
	\begin{equation}
		\begin{cases}
			x_{n+1}=1+y_n-ax^2_n, \\
			y_{n+1}=bx_n,
		\end{cases}
		\label{eq:1}
	\end{equation}
	where $x$ and $y$ represent the outputs of the Hénon map, and $a$ and $b$ represent the control parameters of the Hénon map. When $a=1.4$ and $b=0.3$, the map enters a chaotic state, and the system exhibits the well-known Hénon attractor (as shown in Fig. ~\ref{fig:1}), with the resulting sequences displaying good randomness and long-term uncertainty \cite{Ping2015}.
	\subsection{Algorithm Model}
	\label{sec2.2}
	In this section, we will introduce the algorithmic model used in this work, namely the SINDy-PI algorithm. We will provide a brief overview of the algorithm, covering its development process, principles, operational logic, and the progress achieved.
	In 2016, researchers first proposed the Sparse Identification of Nonlinear Dynamics (SINDy) algorithm\cite{Brunton2016}. By constructing a basis function library and combining it with L1 regularization optimization, this algorithm represents system dynamics as a linear combination of a few key terms, overcoming the limitations of traditional models that rely on prior assumptions. This algorithm has been widely applied in the study of complex nonlinear systems, enabling the extraction of key dynamical features from large datasets to establish effective dynamical models. To address the shortcomings of the traditional SINDy algorithm in strongly coupled implicit systems, prior to the introduction of SINDy-PI, Implicit SINDy\cite{Mangan2016} had already been developed as an improvement over the original SINDy to identify the implicit equation:
	\begin{equation}
		f(x,\dot x)=0,
		\label{eq:2}
	\end{equation}
	it modifies the trajectory evaluation library proposed in the original SINDy to include both $x$ and $\dot x$, as shown in Eq.~\ref{eq:3}:
	\begin{equation}
		\boldsymbol{\Theta}(\mathbf{X}, \dot{\mathbf{X}}) \boldsymbol{\Xi}=\boldsymbol{0}.
		\label{eq:3}
	\end{equation}
		However, this method suffers from poor noise robustness\cite{Mangan2016}. Therefore, in 2020, researchers further developed the Sparse Identification of Nonlinear Dynamics-Parallel Implicit (SINDy-PI) algorithm\cite{Kaheman2020}. A key feature of this algorithm is that, for the implicit differential equations to be identified, as long as one relationship $\theta_j(x,\dot{x})\in\boldsymbol{\Theta}(x, \dot{x})$ is known in Eq.~\ref{eq:2}, we can rewrite Eq.~\ref{eq:3} as:
	\begin{equation}
		\theta_j(\mathbf{X},\dot{\mathbf{X}})=\boldsymbol{\Theta}(\mathbf{X},\dot{\mathbf{X}}|\theta_j(\mathbf{X},\dot{\mathbf{X}}))\boldsymbol{\xi}_j,
		\label{eq:4}
	\end{equation}
	where $\boldsymbol{\Theta}(\mathbf{X},\dot{\mathbf{X}}|\theta_j(\mathbf{X},\ \dot{\mathbf{X}}))$ is the part of the library $\boldsymbol{\Theta}(\mathbf{X}, \dot{\mathbf{X}})$ corresponding to $\theta_j$, and $\boldsymbol{\xi}_j$ is a sparse coefficient vector. At this point, Eq.~\ref{eq:4} is no longer in implicit form\cite{Kaheman2020}. In the algorithm, $\boldsymbol{\xi}_j$ is solved to minimize the following loss function:
	\begin{equation}
		\| \theta_{j}(\mathbf{X}, \dot{\mathbf{X}})-\boldsymbol{\Theta}\left(\mathbf{X}, \dot{\mathbf{X}} \mid \theta_{j}(\mathbf{X}, \dot{\mathbf{X}}) \boldsymbol{\xi}_{j}\left\|_{2}+\beta\right\| \boldsymbol{\xi}_{j} \|_{0}\right.,
		\label{eq:5}
	\end{equation}
	where $\beta$ is used to promote sparsity of the parameters.
	
	The algorithm selects candidate models from a library of candidate functions; when the candidate function $\theta_j$ does not appear in the true dynamical equations, the resulting coefficient vector $\boldsymbol{\xi}_j$ will not be sparse. Iterative calculations are performed using Sequential Threshold Least Squares (STLSQ) to minimize $\left\|\theta_{j}(\mathbf{X}, \dot{\mathbf{X}})-\boldsymbol{\Theta}\left(\mathbf{X}, \dot{\mathbf{X}} \mid \theta_{j}(\mathbf{X}, \dot{\mathbf{X}})\right) \boldsymbol{\xi}_{j}\right\|_{2}$, and retaining terms of $\boldsymbol{\xi}_j$ that are below a preset threshold $\lambda$. Scores are calculated for all obtained model function terms belonging to the true dynamical equations, and the optimal expression is derived and output \cite{Kaheman2020}.
	
	To date, the effectiveness of the SINDy-PI algorithm has been validated across multiple fields and has yielded numerous research results. For example, upon the algorithm’s inception, researchers demonstrated its strong robustness in identifying double-pendulum dynamics and the Belousov–Zhabotinsky (BZ) reaction\cite{Kaheman2020}; Reference \cite{Mangan2016} has demonstrated the immense research value of sparse nonlinear dynamics in biological networks, and SINDy-PI has also shown great potential in studying the behavior of complex biological systems \cite{Massonis2023}, among other findings. It is worth noting that the above studies have all focused on the algorithm’s application to continuous-time systems, while there has been little discussion regarding discrete-time systems.
	Given that the development of cryptographic algorithms is always based on discrete-time systems, this work improves the algorithm through the following steps: First, the library of preselection functions is reconstructed based on mapping characteristics to ensure it includes the required elements; second, the data generation logic is adjusted so that methods such as the fourth-order Runge-Kutta method are no longer used to generate datasets based on time steps.
	
	It is worth noting that improvements to the algorithm have revealed that it can be used for the identification of discrete-time systems with high accuracy and robustness. This breakthrough provides a foundation for the optimization and further research of the encryption algorithm discussed below; specific results will be presented and explained in the following sections.

	\section{Data and Exploration}\label{sec3}
	\subsection{Data Generation Method Correction}
	\label{sec3.1}
		In the SINDy-PI source code, data is generated using numerical methods such as the fourth-order Runge-Kutta method, which depends on the time step and total duration. Since the chaotic systems studied in this paper are all discrete-time systems, this approach is no longer applicable. To address this, we have adopted a cyclic iteration method: based on the required data size $n$, the initial state is repeatedly substituted into the system equations to iteratively generate and store the sequence. This modification alters only the data generation logic without changing the function’s inputs, outputs, or calling interface. It achieves a smooth transition from continuous to discrete while preserving the code structure, and facilitates the subsequent addition of features such as noise generation and randomization. Experiments in the following sections confirm that this approach is correct and effective.
	
	\subsection{Data Generation and Results}
	\label{sec3.2}
	In data-driven modeling, the accuracy of the data directly affects the algorithm’s recognition results. Building on the improvements to data generation methods discussed earlier, this section conducts a preliminary validation using a smaller dataset. The significance threshold for coefficients in the algorithm is set to 0.0001; terms below this threshold are deemed insufficient to characterize the system’s features and are discarded. This mechanism enhances the model’s interpretability during dynamic optimization.
	Taking the classic Hénon map as an example, 10,000 data sets were generated iteratively using the initial conditions (Eq.~\ref{eq:6}), and the learning results are shown in Table~\ref{table1}.
	\begin{equation}
		\begin{cases}
			x=0.1,\\
			y=0.1.
		\end{cases}
		\label{eq:6}
	\end{equation}

\begin{table}[htbp]
	\centering
	\caption{Learning results for the classical Hénon map.}
	\label{table1}
	\begin{tabular*}{\linewidth}{@{\extracolsep{\fill}}Lr@{}}
		\toprule
		\textbf{Standard Equations} 
		& $\begin{aligned}
			x_{n+1} &= 1-1.4x_{n}^2+y_n \\
			y_{n+1} &= 0.3x_{n}
		\end{aligned}$ \\
		\midrule
		\textbf{Noise = 0} & \\
		\textbf{Data size = 10000} 
		& $\begin{aligned}
			x_{n+1} &= 1-1.4x_{n}^2+y_n \\
			y_{n+1} &= 0.3x_{n}
		\end{aligned}$ \\
		\bottomrule
	\end{tabular*}
\end{table}
	
	Under noise-free conditions, the algorithm accurately identifies the two-dimensional Hénon map. However, since this conclusion is based on a fixed dataset, we will investigate the impact of varying dataset sizes on recognition accuracy in future work.
	\subsection{Error Analysis}
	\label{sec3.3}
	To investigate the effect of data size on recognition results, this section generates 10 data sets by incrementing the data size from 2,000 to 20,000 in increments of 2,000, and quantitatively evaluates the learning performance using an error function. The error function is defined as in Eq.\ref{eq:7}:
	\begin{equation}
		error=\sqrt{(A-a)^2+(B-b)^2+(C-c)^2+\cdots},
		\label{eq:7}
	\end{equation}
	where $A, B, C$ are the control parameters of the system model we input, and $a, b, c$ are the system parameters learned by the algorithm. This function characterizes the quality of the learning performance by comparing the algorithm’s learning results with the true system parameters under different data sizes.
	
	When there is a discrepancy between the algorithm’s output coefficients and the true coefficients, we calculate the difference between them, square it, and add it to the error function. Similarly, when extraneous disturbance terms appear in the algorithm’s results, we directly square their coefficients and add them to the error function.
	
\begin{figure}[h]
		\centering
		\includegraphics[width=0.5\textwidth]{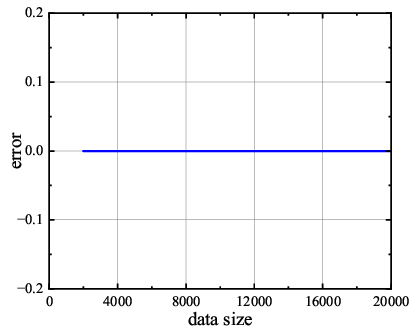}
		\caption{Error plot of the classical Hénon map}
		\label{fig:2}
	\end{figure}

	The curve showing the system error as a function of data size is shown in Fig.~\ref{fig:2}.
	As shown in Fig. ~\ref{fig:2}, as the amount of data gradually increases, the error value $error$ remains constant at 0, indicating that the algorithm’s learning ability does not decline. In other words, the algorithm’s recognition accuracy for this system is independent of the amount of data.
	\subsection{Noise Robustness}
	\label{sec3.4}
		To investigate the algorithm’s robustness under noisy conditions, we add Gaussian white noise to data generated by the two-dimensional Hénon map.
	In short, for the original noise-free data $X$, the noisy data $X_{\text{noisy}}$ is generated by Eq.~\ref{eq:8}:
	\begin{equation}
		X_{\text{noisy}} = X + \sigma \cdot \mathcal{N}(0,1),
		\label{eq:8}
	\end{equation}
	where $\mathcal{N}(0,1)$ denotes the standard normal distribution, and $\sigma$ is the noise intensity (standard deviation).
	
\begin{figure}[h]
		\centering
		\includegraphics[width=0.5\textwidth]{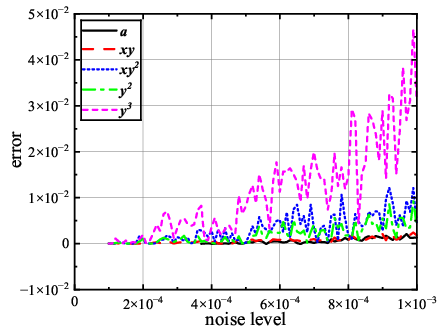}
		\caption{The Effect of Noise on Recognition Error.The black solid line represents the constant term ($a$), the red dashed line represents the interaction term ($xy$), the blue dotted line corresponds to the term ($xy^2$), the green dash-dotted line indicates the quadratic term ($y^2$), and the magenta dash-dotted line shows the cubic term ($y^3$).}
		\label{fig:3}
	\end{figure}

	To evaluate the noise robustness of the SINDy-PI algorithm, we first tested its stability under a noise level of 0.001. The output results contained a large number of redundant terms; however, when the noise level was reduced to 0.0001, the algorithm successfully identified the correct equation. The results are shown in Table~\ref{table2}. To investigate the algorithm’s sensitivity to system noise, we scanned the \(noise \ level\) from \(10^{-4}\) to \(10^{-3}\) in steps of \(10^{-5}\). The results are shown in Fig. ~\ref{fig:3}.
	
	In the figure, $a$ represents the sum of the absolute differences between the coefficients of the Henon map’s eigenvector terms (the constant term, $x_n^2$, $x_n$, and $y_n$) in the equation output by the algorithm and the corresponding terms in the standard Henon map.
	
	Based on the information in the figure, it is clear that as the noise level gradually increases, artifacts begin to appear in the recognition results, and recognition errors show an overall upward trend. Additionally, deviations occur in the system’s original feature coefficients. This indicates that the increase in noise level not only leads to the emergence of redundant terms but also affects the system’s structural recognition.
	\begin{table}[t]
		\centering
		\caption{Results of noise interference tests on the classical Hénon map.}
		\label{table2}
		\begin{tabular*}{\tblwidth}{@{\extracolsep{\fill}} c c L @{}}
			\toprule
			\textbf{Noise} & \textbf{Data} & \multicolumn{1}{l}{\textbf{Learned Equations}} \\
			\midrule
			0 & 10000 & 
			$\begin{aligned}
				x_{n+1} &= 1.0 - 1.4x_n^2 + y_n \\
				y_{n+1} &= 0.3x_n
			\end{aligned}$ \\
			\midrule
			0.0001 & 10000 & 
			$\begin{aligned}
				x_{n+1} &= 1.0 - 1.4x_n^2 + y_n \\
				y_{n+1} &= 0.3x_n
			\end{aligned}$ \\
			\midrule
			0.001 & 10000 & 
			$\begin{aligned}
				x_{n+1} &= 1.0 - 1.4x_n^2 + 1.001y_n - 0.02004y_n^3 - 0.0005039x_n \\
				&\phantom{=} - 0.003496y_n^2 + 0.004254x_ny_n^2 - 0.001436x_ny_n \\
				y_{n+1} &= 0.3x_n + 0.0001284y_n^3
			\end{aligned}$ \\
			\bottomrule
		\end{tabular*}
	\end{table}
	It is worth noting that the pre-selected function library used in the above experiments is sufficiently comprehensive, and the algorithm filters out valid terms based on data correlations. This implies that, under ideal conditions where the pre-selected function library is comprehensive and noise is minimal, this method can be extended to the discovery of dynamical equations for higher-dimensional unknown discrete systems.
	\section{Overview of Image Encryption Schemes}
	\label{sec4}
	Building on the previous modifications to the SINDy-PI algorithm to adapt it to discrete systems, this section explores the feasibility of combining the SINDy-PI algorithm with image encryption using the two-dimensional Henon map as an example, and proposes a viable scheme. The core idea of this scheme is the “key + mapping” approach: during encryption, not only is the key derived from the system's initial state $(x_0, y_0)$ used, but a pseudorandom sequence is also generated via a chaotic mapping to scramble and diffuse the image, ultimately yielding the ciphertext image \cite{Lone2024}, The mapping function is also incorporated into the encryption and decryption process; therefore, during decryption, not only must the same key used for encryption be employed, but the mapping function must also remain consistent to successfully decrypt the image.
	\subsection{key structure}
	\label{sec4.1}
	Unlike traditional encryption schemes, the chaotic map in this scheme is not a predefined fixed form, but rather a complete dynamical equation automatically discovered by SINDy-PI from the training data. Therefore, the key consists solely of the initial state of the chaotic map:
	\begin{equation}
		\mathbf{K} = [x_0,\; y_0]^{\mathrm{T}},
		\label{eq:9}
	\end{equation}
	where \((x_0, y_0)\) is the initial state of the chaotic map. The specific form of the map (e.g., the standard Hénon form or a variant with minor noise) can be made public as part of the encryption algorithm, but since it depends on specific training data, an attacker cannot know it in advance. This design introduces an additional security factor while maintaining the algorithm’s transparency: even if an attacker obtains the key, they cannot correctly decrypt the data without knowing the exact form of the mapping (e.g., whether redundant terms exist).
	\subsection{Chaotic Sequence Generation}
	\label{sec4.2}
		In the experiments described in this paper, the standard Hénon map identified by SINDy-PI from noise-free data is used as an example; when the training data contains noise, the identified map may contain minor artifacts (as shown in Table~\ref{table2}). In such cases, both encryption and decryption use the complete map—including these artifacts—to ensure consistency. Taking the noise-free case as an example, given a key $K$, the Hénon map identified by the algorithm generates a sequence of chaotic states through iteration:
	\begin{equation}
		\begin{cases}
			x_{n+1}=1+y_n-ax^2_n, \\
			y_{n+1}=bx_n.
		\end{cases}
		\label{eq:10}
	\end{equation}
	
	Assume the image size is $M\times N$, and the total number of pixels is $P=MN$. To ensure a sufficiently long sequence for shuffling and diffusion, we iteratively generate a sequence of states of total length $L=M+N+2P+500$, denoted as ${(x_i,y_i)}_{i=1}^L$. The sequence is divided and used as follows:
	
	First, row scrambling: Take the $x$ components of the first $M$ states, i.e., $R = [x_1, x_2, ..., x_M]$.
	
	Second, column scrambling: Take the $y$ components of the first $N$ states, i.e., $C = [y_1, y_2, ..., y_N]$.
	
	Finally, diffusion: take the $x$ components of the next $P$ states as the first-round diffusion sequence $Q_1$, and then take the $x$ components of the subsequent $P$ states as the second-round diffusion sequence $Q_2$ (if multi-round diffusion is used, proceed in this manner successively)\cite{Singh2023}.
	\subsection{A Brief Overview of the Scrambling Algorithm}
	\label{sec4.3}
	The shuffling process uses the classic row-column independent sorting shuffling method:
	First, sort the row shuffling sequence $R$ to obtain the index mapping $row_{idx}$, such that $R(row_{idx})$ is in ascending order. Similarly, sort the column shuffling sequence $C$ to obtain $col_{idx}$. Finally, rearrange the original image $I$ by $row_{idx}$, then rearrange the columns by $col_{idx}$ to obtain the shuffled image $S$:
	\begin{equation}
		S(i,j)=I(row_{idx}(i),col_{idx}(j)).
		\label{eq:11}
	\end{equation}
	This scrambling method is simple and efficient, and due to the randomness of the chaotic sequence, the pixel positions in the scrambled image are uniformly distributed\cite{Bezerra2023}.
	\subsection{Overview of Diffusion Algorithms}
	\label{sec4.4}
	To enhance plaintext sensitivity, multi-round diffusion based on modulo-256 addition is employed. Additive diffusion offers a better avalanche effect than XOR diffusion, ensuring that a single-bit change propagates to all pixels. The quantized chaotic sequence is defined as:
	\begin{equation}
		q_k = \lfloor \operatorname{mod}(|\xi_k|, 1) \times 256 \rfloor, k=1,2,\dots,
		\label{eq:12}
	\end{equation}
	where $\xi_k$ is the original chaotic state component (the fractional part is uniformly mapped to [0, 255]). During encryption, the scrambled image $S$ is flattened into a one-dimensional vector $p$, with the initial vector $iv=0$. The two-round diffusion process (forward + reverse) is as follows:
	
	Forward diffusion:
	\begin{equation}
		t_1(i)=(p(i)+q_1(i)+iv) \pmod{256},
		\label{eq:13}
	\end{equation}
	and the subsequent pixel $iv$ is updated to $t_1(i)$.
	
	Backward diffusion:
	\begin{equation}
		t_2(i)=(t_1(i)+q_2(i)+t_2(i+1)) \pmod{256},
		\label{eq:14}
	\end{equation}
	Working backward from the last pixel, $t_2(P+1)=0$.
	
	The final ciphertext vector $c = t_2$ is reconstructed as an $M \times N$ matrix to obtain the encrypted image $C$. To further enhance the diffusion strength, three or four rounds (forward–backward–forward–backward) can be employed; this simply requires adding corresponding chaotic sequences for each segment. This paper actually adopts four rounds of diffusion to ensure that the plaintext sensitivity reaches the theoretical optimum.
	\subsection{Decryption Process}
	\label{sec4.5}
	During decryption, the decryption end must use the exact same chaotic map (i.e., the same symbolic expression) and the same initial state \((x_0, y_0)\) as the encryption end. Since the map form has been shared in advance (or transmitted via a secure channel) as part of the algorithm, the decryption process is as follows:
	
	First, generate an identical chaotic sequence using the key $K$, and extract the corresponding $Q_1$, $Q_2$, and scrambling indices $row_{idx}$, $col_{idx}$ as used during encryption. Second, perform reverse de-diffusion on the ciphertext vector $c$: 
	\begin{equation}
		t_1(i)=(c(i)-q_2(i)-t_1(i+1)) \pmod{256},
		\label{eq:15}
	\end{equation}
	Calculating from the end to the beginning, $t_1(P+1)=0$. Then perform forward diffusion:
	\begin{equation}
		p(i)=(t_1(i)-q_1(i)-p(i-1)) \pmod{256},
		\label{eq:16}
	\end{equation}
	Computing backward, $p(0)=0$. After restoring the scrambling vector $p$, reconstruct the image $S$. Finally, use the inverse mapping of $row_{idx}$ and $col_{idx}$ to recover the original image:
	\begin{equation}
		I_{dec}(row_{idx}(i),col_{idx}(j))=S(i,j).
		\label{eq:17}
	\end{equation}
	\section{Experimental Results}
	\label{sec5}
\begin{figure*}[htbp]
		\centering
		\begin{minipage}[b]{0.32\linewidth}
			\centering
			\includegraphics[width=\textwidth]{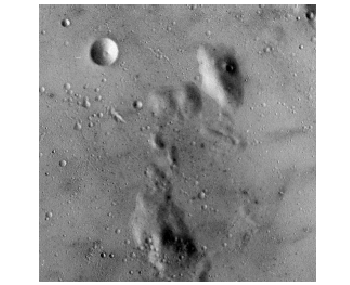}
			\par\vspace{0.5em}
			\textbf{(a)}
		\end{minipage}
		\hfill
		\begin{minipage}[b]{0.32\linewidth}
			\centering
			\includegraphics[width=\textwidth]{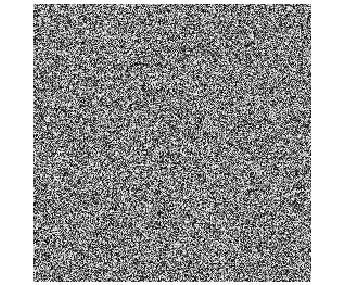}
			\par\vspace{0.5em}
			\textbf{(b)}
		\end{minipage}
		\hfill
		\begin{minipage}[b]{0.32\linewidth}
			\centering
			\includegraphics[width=\textwidth]{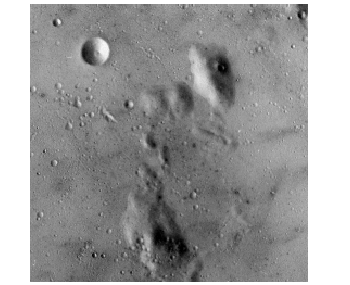}
			\par\vspace{0.5em}
			\textbf{(c)}
		\end{minipage}
		\vspace{0.5em}
	    \caption{Encryption and decryption results: (a) Original image of the Moon surface; (b) Encrypted image of the Moon surface; (c) Decrypted image of the Moon surface.}
	\label{fig:4}
	\end{figure*}

	This section presents a comprehensive experimental evaluation of the proposed data-driven image encryption framework, demonstrating the feasibility of combining the SINDy-PI algorithm with image encryption. It should be noted that this section continues to use the two-dimensional Hénon map as an example, and all experiments employ the $256\times256$ 8-bit grayscale image "Moon surface" as the test sample. The chaotic map used for encryption is the complete equation of SINDy-PI identified from noise-free observational data (the standard Hénon map $x_{n+1}=1-1.4x_n^2+y_n,\ y_{n+1}=0.3x_n$), with the key consisting solely of the initial state $[x_0, y_0] = [0.2, 0.3]$. 
	The original image, the​ encrypted image, and the decrypted image are shown in Fig.~\ref{fig:4}, respectively;​ PSNR is infinite (lossless decryption).
	The following sections will provide a detailed explanation of the effectiveness and versatility of this encryption scheme from eight perspectives.
	
	\subsection{Key Space}
	\label{sec5.1}
	In this scheme, the key consists solely of two double-precision floating-point initial values \((x_0, y_0)\), resulting in an effective key space of approximately $(10^{15})^{2}=10^{30} \approx 2^{100}$. This space is sufficient to defend against any brute-force attack\cite{Chai2020,Yang2009}. Although this value is the same as that of fixed-mapping schemes (which rely solely on initial values), the security of this scheme additionally depends on the data dependency of the mapping form. If an attacker does not know the exact form of the mapping (e.g., whether there are minor random terms), they cannot correctly decrypt the data even if they obtain the initial values. Therefore, the actual difficulty of breaking the system is far greater than the complexity implied by the key space value. Furthermore, by altering the training data (e.g., using different noise seeds), entirely different mapping forms can be obtained, thereby supporting a “one-time pad” variant encryption strategy.The impact of mapping structures on encryption and decryption will be discussed in Sec.~\ref{sec5.7}
	\subsection{Key Sensitivity}
	\label{sec5.2}
	High key sensitivity is a key characteristic of any encryption algorithm. In other words, even slight variations in the encryption and decryption keys can lead to significant differences in experimental results. In this paper, we set the initial values of the two-dimensional Hénon map, $x_0=0.2$ and $y_0=0.3$, as the correct key to encrypt the grayscale image "Moon surface ($256\times256$)."
	
	In this experiment, we applied small perturbations ranging from \(10^{-16}\) to \(10^{-12}\) to the two initial components \(x_0\) and \(y_0\) of the key, and then used the perturbed key to decrypt the original ciphertext. As shown in Table~\ref{table3}, the PSNR of the decrypted images remains stable at around 10.2 dB for all perturbation magnitudes, which is far below the lossless threshold of 30 dB. This indicates that even a perturbation of \(10^{-16}\) is sufficient to completely compromise decryption. This demonstrates that the algorithm is extremely sensitive to the initial values.
	\begin{table}[htbp]
		\centering
		\caption{Key sensitivity evaluation via PSNR (dB).}
		\label{table3}
		\begin{tabular*}{\tblwidth}{@{\extracolsep{\fill}}ccc@{}}
			\toprule
			\textbf{Perturbation Magnitude} & \textbf{$x_0$} & \textbf{$y_0$} \\
			\midrule
			2.782559e-16 & 10.21 & 10.16 \\
			7.742637e-15 & 10.20 & 10.19 \\
			2.154435e-15 & 10.16 & 10.18 \\
			5.994843e-15 & 10.18 & 10.14 \\
			1.668101e-14 & 10.21 & 10.19 \\
			4.641589e-14 & 10.20 & 10.22 \\
			1.291550e-13 & 10.18 & 10.19 \\
			3.593814e-13 & 10.19 & 10.17 \\
			1.000000e-12 & 10.18 & 10.19 \\
			\bottomrule
		\end{tabular*}
		\par\vspace{0.5em}
		\footnotesize\textit{Note:} Perturbations below $2\times10^{-16}$ were ineffective due to floating-point precision limits.
	\end{table}
	
	The Normalized Pixel Change Rate (NPCR) and the Unified Average Change Intensity (UACI) are two quantitative tests for verifying the sensitivity of encryption keys\cite{Qian2021,Lone2024,Alawida2019}.
	For two ciphertext images $\mathbf{C}_1$ and $\mathbf{C}_2$ of size $M\times N$, NPCR and UACI are defined as follows:
	
	\begin{equation}
		\label{eq:18}
		\mathrm{NPCR} = \frac{1}{M\times N}\sum_{i=1}^{M}\sum_{j=1}^{N} D(i,j) \times 100\%,
	\end{equation}
	where $D(i,j) = 1$ if $\mathbf{C}_1(i,j) \neq \mathbf{C}_2(i,j)$, otherwise $D(i,j) = 0$.
	
	\begin{equation}
		\label{eq:19}
		\mathrm{UACI} = \frac{\sum_{i=1}^{M}\sum_{j=1}^{N} \bigl|\mathbf{C}_1(i,j) - \mathbf{C}_2(i,j)\bigr| \times 100\%}{M\times N \times 255}.
	\end{equation}
	
	For two ideal, independent, and identically distributed images, the expected values of NPCR and UACI are:
	
	\begin{equation}
		\label{eq:20}
		\mathrm{NPCR}  \approx 99.6094\%,\quad
		\mathrm{UACI}  \approx 33.4635\%.
	\end{equation}
	
	A secure encryption algorithm should produce NPCR and UACI values that are very close to these theoretical ideal values, indicating that even the slightest variation in the key or plaintext is effectively spread across the entire ciphertext image.
	
	\begin{table}[htbp]
		\centering
		\caption{Key Sensitivity: NPCR and UACI VS Perturbation Magnitude.}
		\label{table4}
		\begin{tabular*}{\tblwidth}{@{\extracolsep{\fill}} c S[table-format=2.4] S[table-format=2.4] S[table-format=2.4] S[table-format=2.4] @{}}
			\toprule
			\multirow{2}{*}{\centering Perturbation Magnitude} & 
			\multicolumn{2}{c}{$x_0$} & 
			\multicolumn{2}{c}{$y_0$} \\
			\cmidrule(lr){2-3} \cmidrule(lr){4-5}
			& {NPCR (\%)} & {UACI (\%)} & {NPCR (\%)} & {UACI (\%)} \\
			\midrule
			2.782559e-16 & 99.6033 & 33.4667 & 99.6033 & 33.4667 \\
			7.742637e-16 & 99.5926 & 33.2886 & 99.5926 & 33.2886 \\
			2.154435e-15 & 99.6277 & 33.4513 & 99.6277 & 33.4513 \\
			5.994843e-15 & 99.6323 & 33.4067 & 99.6323 & 33.4067 \\
			1.668101e-14 & 99.6109 & 33.5000 & 99.6109 & 33.5000 \\
			4.641589e-14 & 99.6063 & 33.5011 & 99.6063 & 33.5011 \\
			1.291550e-13 & 99.6033 & 33.4774 & 99.6033 & 33.4774 \\
			3.593814e-13 & 99.6307 & 33.5446 & 99.6307 & 33.5446 \\
			1.000000e-12 & 99.5819 & 33.4543 & 99.5819 & 33.4543 \\
			\bottomrule
		\end{tabular*}
	\end{table}
	Table~\ref{table4} shows that the algorithm developed in this work is highly sensitive to the key. Even when the perturbation is minimal, it can still be effectively propagated throughout the ciphertext image, leading to decryption failure.
	\subsection{Histogram Analysis}
	\label{sec5.3}
\begin{figure*}[htbp]
		\centering
		\begin{subfigure}[b]{0.45\textwidth}
			\centering
			\includegraphics[width=\textwidth]{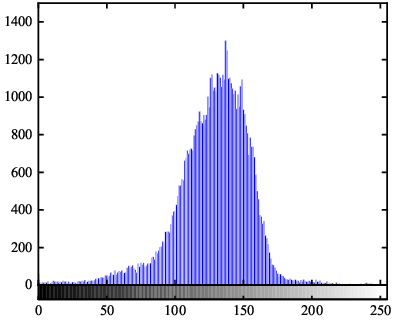}
			\label{fig:5a}
		\end{subfigure}
		\hfill
		\begin{subfigure}[b]{0.45\textwidth}
			\centering
			\includegraphics[width=\textwidth]{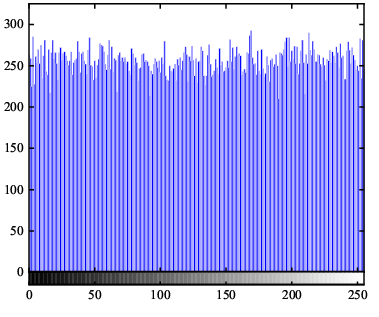}
			\label{fig:5b}
		\end{subfigure}
		\caption{Histogram: (a) Histogram of the original image; (b) Histogram of the encrypted image.}
		\label{fig:5}
	\end{figure*}

	A histogram provides a visual representation of the grayscale distribution of an image’s pixels. Fig. ~\ref{fig:5} shows the histograms of the original image and the encrypted image. The histogram of the original image exhibits a distinctly non-uniform distribution, reflecting a strong correlation among the pixels; in contrast, the histogram of the encrypted image approximates a horizontal line, indicating that the pixel values are uniformly distributed across all grayscale levels, with no exploitable statistical features.
	
		To further quantitatively evaluate the uniformity of the encrypted image, a chi-square test is used to analyze the grayscale histogram of the encrypted image\cite{Huang2015}. For a grayscale image of size $M\times N$, under an ideal uniform distribution, the expected frequency of each grayscale level is $E = \frac{(M\times N)}{256}$.In a $256\times256$ image, its value is 256. The chi-square statistic is defined as:
	\begin{equation}
		\chi^2 = \sum_{i=0}^{255} \frac{(O_i - E)^2}{E},
		\label{eq:21}
	\end{equation}
	where $O_i$ denotes the observed pixel count at gray level $i$. With $256$ gray levels, the degrees of freedom are $\nu = 255$. At a significance level $\alpha = 0.05$, the critical value is $\chi^2_{0.05}(255) \approx 293.25$. If the computed $\chi^2$ is less than this critical value, the null hypothesis of uniformity is not rejected; conversely, a larger $\chi^2$ would indicate a statistically significant deviation from uniformity.

	\begin{table}[htbp]
		\centering
		\caption{Chi-square goodness-of-fit test for histograms of original and encrypted images.}
		\label{table5}
		\begin{tabular*}{\linewidth}{@{\extracolsep{\fill}}lcccc@{}}
			\toprule
			\textbf{Image} & \textbf{$\chi^2$ Statistic} & \textbf{$p$-value} & \textbf{Critical Value ($\alpha=0.05$)} & \textbf{Decision} \\
			\midrule
			Original image & 135687.5703 & 0.0000 & 293.25 & Reject \\
			Encrypted image & 212.7969 & 0.9746 & 293.25 & Pass \\
			\bottomrule
		\end{tabular*}
	\end{table}

	The obtained chi-square statistics are $\chi^2 = 135687.5703$ ($p=0.0000$) for the original image and $\chi^2 = 212.7969$ ($p=0.9746$) for the ciphertext image (see Table~\ref{table5}). For the original image, $\chi^2$ far exceeds the critical value $293.25$ and $p<0.05$, indicating a significant deviation from a uniform distribution – as expected for natural images. In contrast, the ciphertext image yields $\chi^2 < 293.25$ and $p>0.05$, so the null hypothesis of uniformity is accepted. Therefore, the grayscale distribution of the encrypted image does not differ significantly from an ideal uniform distribution. This result confirms that the proposed encryption scheme effectively eliminates statistical patterns, demonstrating strong resistance to histogram-based attacks.
	\subsection{Correlation Analysis}
	\label{sec5.4}
\begin{figure*}[htbp]
		\centering
		
		\begin{subfigure}[b]{0.48\textwidth}
			\centering
			\includegraphics[width=\textwidth]{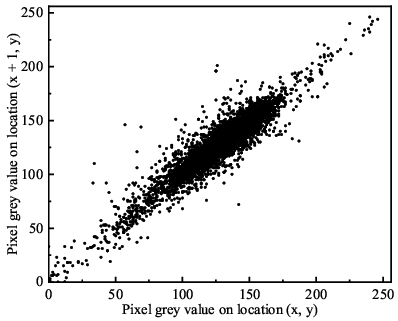}
			\label{fig:6a}
		\end{subfigure}
		\hfill
		\begin{subfigure}[b]{0.48\textwidth}
			\centering
			\includegraphics[width=\textwidth]{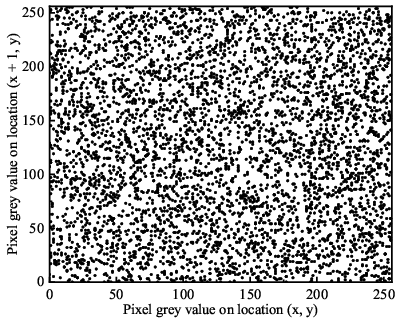}
			\label{fig:6b}
		\end{subfigure}
		
		\vspace{1em}
		
		\begin{subfigure}[b]{0.48\textwidth}
			\centering
			\includegraphics[width=\textwidth]{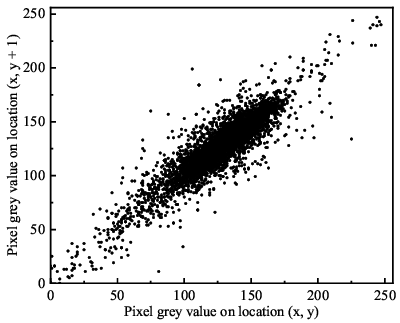}
			\label{fig:6c}
		\end{subfigure}
		\hfill
		\begin{subfigure}[b]{0.48\textwidth}
			\centering
			\includegraphics[width=\textwidth]{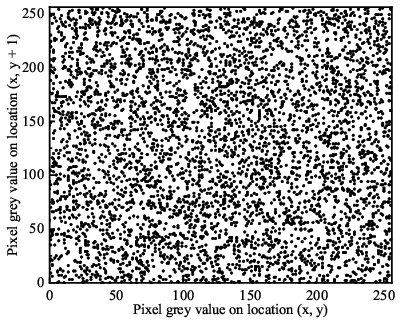}
			\label{fig:6d}
		\end{subfigure}
		
		\vspace{1em}
		
		\begin{subfigure}[b]{0.48\textwidth}
			\centering
			\includegraphics[width=\textwidth]{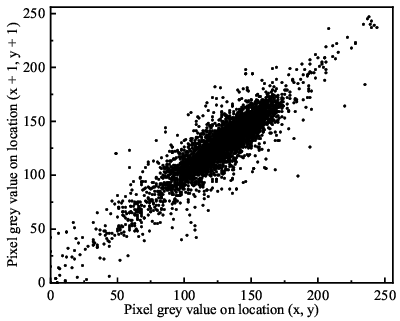}
			\label{fig:6e}
		\end{subfigure}
		\hfill
		\begin{subfigure}[b]{0.48\textwidth}
			\centering
			\includegraphics[width=\textwidth]{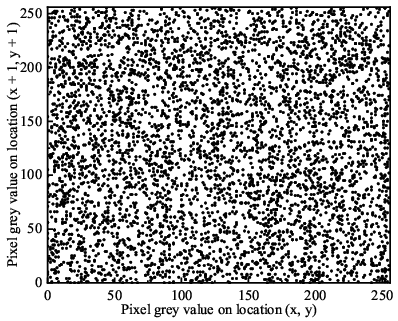}
			\label{fig:6f}
		\end{subfigure}
		
		\caption{Distribution of adjacent pixel correlations: (a) Plaintext (Horizontal); (b) Ciphertext (Horizontal); (c) Plaintext (Vertical); (d) Ciphertext (Vertical); (e) Plaintext (Diagonal); (f) Ciphertext (Diagonal).}
		\label{fig:6}
	\end{figure*}

	In the security evaluation of image encryption algorithms, the correlation between adjacent pixels is a key metric for assessing the algorithm’s scrambling effectiveness. In natural images, adjacent pixels often exhibit a high degree of correlation (with a correlation coefficient close to 1), whereas an ideal encryption algorithm should be able to completely destroy this correlation, making any pair of adjacent pixels in the encrypted image approximately independent, with the correlation coefficient approaching zero. To measure the correlation between the original image and its corresponding encrypted image, 5,000 pairs of adjacent pixels were randomly selected in three directions, and their correlation coefficients were calculated and compared. The experimental results are shown in Fig. ~\ref{fig:6}. The correlation between adjacent pixels in the original image is very high, whereas the correlation between adjacent pixels in the encrypted image image is significantly reduced.
	
	In particular, the correlation coefficient $r_{xy}$ is defined as:
	\begin{equation}
		\label{eq:22}
		r_{xy} = \frac{\operatorname{cov}(x,y)}{\sqrt{D(x)}\sqrt{D(y)}},
	\end{equation}
	where,
	\begin{align}
		\operatorname{cov}(x,y) &= \frac{1}{P}\sum_{i=1}^{P} \bigl(x_i - E(x)\bigr)\bigl(y_i - E(y)\bigr), \label{eq:23}\\
		D(x) &= \frac{1}{P}\sum_{i=1}^{P} \bigl(x_i - E(x)\bigr)^2, \label{eq:24}\\
		E(x) &= \frac{1}{P}\sum_{i=1}^{P} x_i, \label{eq:25}
	\end{align}
	where $x_i$ and $y_i$ denote the gray-scale values of the first and second pixels in the $i$-th pair of adjacent pixels, respectively, and $P$ is the number of randomly selected pixel pairs (in this paper, $P=5000$). Ideally, the correlation coefficient of the encrypted image should be close to $0$, indicating no linear dependence between pixels~.
	
	Table~\ref{table6} presents and compares the correlation coefficients of adjacent pixels in the horizontal, vertical, and diagonal directions for the plaintext image, the image encrypted using the proposed scheme, and the image encrypted using a fixed-parameter scheme.It can be seen that, after encryption, the correlation between pixels in the proposed scheme (which employs the chaotic mapping discovered from the data using SINDy-PI) is not only close to 0 but also lower than that of traditional encryption schemes with fixed initial values, with a reduction of approximately 15\% in all three directions. This result indicates that the proposed data-driven encryption framework effectively breaks the linear correlation between pixels in the original image, demonstrating a high level of resistance to statistical attacks.
	\begin{table}[htbp]
		\centering
		\caption{Correlation coefficients of adjacent pixels.}
		\label{table6}
		\begin{tabular*}{\tblwidth}{@{\extracolsep{\fill}}lccc@{}}
			\toprule
			\textbf{Image Type} & \textbf{Horizontal} & \textbf{Vertical} & \textbf{Diagonal} \\
			\midrule
			Plaintext            & 0.9432 & 0.9678 & 0.9221 \\
			Fixed-parameter Scheme & 0.0971 & 0.0971 & 0.0971 \\
			Proposed Scheme     & 0.0818 & 0.0825 & 0.0825 \\
			\bottomrule
		\end{tabular*}
	\end{table}
	
	\subsection{Information Entropy}
	\label{sec5.5}
	Information entropy is an important measure of the randomness in the grayscale distribution of image pixels\cite{Wang2019}. For an 8-bit grayscale image with $L=256$ grayscale levels, its information entropy $H$ is defined as:
	\begin{equation}
		\label{eq:26}
		H = -\sum_{i=0}^{255} p(i) \log_2 p(i),
	\end{equation}
	where $p(i)$ is the probability of the gray level $i$ occurring. When all gray levels are equally likely, the information entropy reaches its maximum value of $8$; therefore, the information entropy of an ideal encrypted image should be as close to $8$ as possible.
	
	Table~\ref{table7} lists the information entropy of the original image, the image encrypted with fixed parameters, and the image encrypted using the proposed scheme. The information entropy of the original image is $6.7093$, which is significantly lower than the ideal value, indicating that its grayscale distribution exhibits marked non-uniformity. The information entropy of the image encrypted by the proposed scheme reaches $7.9976$, which is very close to $8$, indicating that the pixel value distribution is highly uniform and lacks exploitable statistical features. Furthermore, the results of the proposed scheme are nearly identical to those of the fixed-parameter scheme (information entropy $7.9972$), proving that using data-driven chaotic mappings for encryption does not reduce the randomness of the encrypted image.
	\begin{table}[htbp]
		\centering
		\caption{Information entropy analysis.}
		\label{table7}
		\begin{tabular*}{\tblwidth}{@{\extracolsep{\fill}}lc@{}}
			\toprule
			\textbf{Image Type} & \textbf{Entropy (bits)} \\
			\midrule
			Plaintext                   & 6.7093 \\
			Fixed-parameter Scheme      & 7.9972 \\
			Proposed Scheme            & 7.9976 \\
			\bottomrule
		\end{tabular*}
	\end{table}
	
	\subsection{Differential Attack}
	\label{sec5.6}
	As a type of chosen-plaintext image attack, differential attacks can uncover correlations between the original input image and the corresponding output image by comparing and analyzing specific differences, while resistance to differential attacks is measured using NPCR and UACI.
	
	To comprehensively evaluate the algorithm’s sensitivity to pixel modifications at arbitrary locations, we conducted multi-location random tests. The specific procedure is as follows: We randomly selected 50 different pixel locations in the original image and flipped the least significant bit (LSB) of the grayscale value at each location, resulting in 50 modified images that differ by only a single bit. We encrypted the original image and each modified image using the same key (containing only the initial value), obtaining corresponding ciphertext image pairs, and calculated the NPCR and UACI for each pair. Finally, we calculated the minimum, maximum, and average values for these 50 tests; the results are listed in Table~\ref{table8}.
	
	As shown in the table, the average NPCR value reaches 99.6556\%, which is very close to the theoretical ideal value of 99.6094\%; the average UACI value is 33.5824\%, which closely matches the theoretical value of 33.4635\%. Regardless of which pixel is selected for modification, the algorithm can propagate a single-bit change across the entire ciphertext image, producing ciphertext differences that approach the level of random variation. Compared with the references ~\cite{Yang2019,Hu2020}, the differential attack resistance of the algorithm in this paper is equally excellent.
	\begin{table}[htbp]
		\centering
		\caption{Differential attack resistance evaluation.}
		\label{table8}
		\begin{tabular*}{\tblwidth}{@{\extracolsep{\fill}}lccc@{}}
			\toprule
			\textbf{Metric} & \textbf{Min (\%)} & \textbf{Max (\%)} & \textbf{Avg (\%)} \\
			\midrule
			NPCR & 99.0265 & 99.9832 & 99.6556 \\
			UACI & 31.1547 & 36.1297 & 33.5824 \\
			\bottomrule
		\end{tabular*}
	\end{table}
	
	\subsection{Map Structure as an Implicit Key}
		\label{sec5.7}
		A distinctive feature of our framework is that the chaotic map itself is derived from the training data. Consequently, even with a fixed initial condition $(x_0,y_0)$, different training data (e.g., with different noise seeds) can yield slightly different map structures. To quantify this extra security layer, we performed the following experiment:
		
		\begin{enumerate}
			\item Fix $(x_0,y_0)=(0.2,0.3)$.
			\item Train SINDy-PI on noise‑free data to obtain $M_{\text{clean}}$ (standard Hénon).
			\item Train SINDy-PI on data with a very low noise level ($\sigma=10^{-4}$) to obtain $M_{\text{noisy}}$, which differs from $M_{\text{clean}}$ by small coefficient perturbations (e.g., $|1.4-1.3999|\approx 10^{-4}$) and may include extra weak cross‑terms.
			\item Encrypt the same plaintext image using $M_{\text{clean}}$ and $M_{\text{noisy}}$ separately, yielding ciphertexts $C_{\text{clean}}$ and $C_{\text{noisy}}$.
			\item Compute NPCR and UACI between $C_{\text{clean}}$ and $C_{\text{noisy}}$.
		\end{enumerate}
		
		The measured NPCR and UACI are $99.6231\%$ and $33.4057\%$, respectively – values extremely close to the theoretical ideals for two independent random images. This demonstrates that even a tiny variation in the map (induced by the training data) results in completely different ciphertexts. Compared with the key sensitivity studied in Sec.~\ref{sec5.2} (which changes the key while keeping the map fixed), the sensitivity to the training data is a unique property of our data‑driven approach. In practice, an attacker who obtains the correct initial condition $(x_0,y_0)$ but does not know the exact map (i.e., the training data) cannot decrypt the ciphertext. Hence, the data‑dependent nature of the discovered map provides an additional implicit key, further enhancing the overall security.
		
		\subsection{Generalization to Other Discrete Maps}
		\label{5.8}
		To demonstrate that our SINDy-PI-based cryptographic framework is not limited to the Hénon map, we also applied it to two other discrete chaotic systems:the three-dimensional logistic map and the two-dimensional piecewise-linear Lozi map. For each system, we generated noise-free training data (after removing transients and performing 10,000 iterations), ran SINDy-PI to recover the mapping structure, and then applied the recovered mapping to the exact same encryption pipeline described in Sec. \ref{sec4} (without any modifications).

		Table~\ref{table9} summarizes the recognition results. In both cases, SINDy-PI recovered the exact functional form with negligible coefficient error (the mean squared error is below the significance threshold of $10^{-4}$). Encrypting the standard test image ``Moon surface'' ($256\times256$ grayscale) using the recovered mappings yields nearly ideal security metrics: information entropies of $\approx 7.9970$ and $\approx 7.9974$ bits, respectively.
		\begin{table}[htb]
			\centering
			\caption{Identification results and encryption performance for different chaotic maps.}
			\label{table9}
			\begin{tabular*}{\tblwidth}{@{\extracolsep{\fill}}lccc@{}}
				\toprule
				\textbf{System} & \textbf{True map} & \textbf{Identified map} & \textbf{Information entropy} \\
				\midrule
				3D Logistic & 
				$\begin{aligned}
					x_{i+1} &= 3.8 x_i(1-x_i) + 0.01 y_i^2 x_i + 0.01 z_i^3 \\
					y_{i+1} &= 3.8 y_i(1-x_i) + 0.01 z_i^2 y_i + 0.01 x_i^3 \\
					z_{i+1} &= 3.8 z_i(1-z_i) + 0.01 x_i^2 z_i + 0.01 y_i^3
				\end{aligned}$ & 
				Same form, error $<10^{-4}$ & 7.9970 \\
				\midrule
				2D Lozi & 
				$\begin{aligned}
					x_{i+1} &= 1 - 1.7|x_i| + y_i \\
					y_{i+1} &= 0.5x_i
				\end{aligned}$ & 
				Same form, error $<10^{-4}$ & 7.9974 \\
				\bottomrule
			\end{tabular*}
		\end{table}
		
		These results confirm that our data-driven encryption framework works effectively across different types of discrete chaotic maps (polynomial, piecewise linear, and one-dimensional), highlighting its general applicability and the robustness of our adaptation of SINDy-PI to discrete-time systems.
		
		\section{CONCLUSION}
		\label{sec6}
		We have demonstrated a data-driven chaotic cryptography framework in which the encryption map is not predefined, but autonomously discovered from data via the SINDy-PI algorithm. The validity of this approach has been verified on three distinct chaotic systems-the classic Hénon map, the three‑dimensional logistic map, and the piecewise-linear Lozi map-confirming its generality across different dynamical structures. More importantly, we experimentally show that even when the initial conditions are fixed, different training data (e.g., with a tiny noise seed) lead to slightly different maps, which produce completely different ciphertexts (NPCR $\approx 99.6\%$, UACI $\approx 33.5\%$). This data‑dependent nature of the discovered map acts as an implicit key, providing an extra security layer beyond the conventional key space. The encryption key consists solely of initial conditions; a perturbation as small as $10^{-16}$ triggers total decryption failure, confirming extreme sensitivity. The cipher achieves near-ideal information entropy ($\approx 8$ bits), vanishing inter-pixel correlation, and full compliance with NPCR and UACI benchmarks, ensuring resistance to statistical and differential attacks. This work establishes a new paradigm for chaos-based cryptography by unifying system identification with encryption, opening a pathway toward robust, adaptive, and data-centric security architectures. Future work includes extending the method to hyperchaotic systems and validating it on experimental time series from physical chaotic circuits.
		
		\printcredits
		
		\section*{Declaration of Competing Interest}
		The authors declare that they have no known competing financial interests or personal relationships that could have appeared to influence the work reported in this paper.
		\section*{Acknowledgements}
		This work is supported by the National Natural Science Foundation of China under Grant No. 12047501, the Natural Science Foundation of Gansu province under Grant Nos. 22JR5RA266, 25JRRA071 and 25JRRA105.
		\section*{Data and code availability}
		The complete source code for the SINDy-PI based identification and the image encryption framework is publicly available at \url{https://github.com/WYLEE2005/SINDy-PI-encrypt}. The repository includes a detailed README with instructions for reproducing all results in this paper.

	\bibliographystyle{elsarticle-num}

	\bibliography{cas-refs}
	
\end{document}